\documentclass[aps,twocolumn,superscriptaddress, showpacs,showkeys,preprintnumbers, amsmath,amssymb,amsfonts,nofootinbib]{revtex4}

\pdfoutput=1

\usepackage{graphicx}
\usepackage{color}

\def\0#1#2{\frac{#1}{#2}}

\def\s0#1#2{\mbox{\small{$ \frac{#1}{#2} $}}}

\newcommand{\Tr}{\mathrm{Tr}}

\newcommand{\be}{\begin{eqnarray}}
\newcommand{\ee}{\end{eqnarray}}

\newcommand{\nn}{\nonumber }

\newcommand{\Nc}{N_{\text{c}}}

\begin{document}

\title{{The Effect of Fluctuations} on the QCD Critical Point in a Finite Volume}

\author{Ralf-Arno Tripolt} \affiliation{Institut f\"ur Kernphysik
  (Theoriezentrum), Technische Universit\"at Darmstadt, Schlo\ss gartenstra\ss e 2, D-64289
  Darmstadt, Germany} \affiliation{Institut f\"ur Physik, Karl-Franzens-Universit\"at {Graz}, A-8010
  Graz, Austria}
     \author{Jens~Braun} \affiliation{Institut f\"ur Kernphysik
  (Theoriezentrum), Technische Universit\"at Darmstadt, Schlo\ss gartenstra\ss e 2, D-64289
  Darmstadt, Germany} \affiliation{ExtreMe Matter Institute EMMI, GSI,
  Planckstra{\ss}e 1, D-64291 Darmstadt, Germany} \affiliation{Theoretisch-Physikalisches Institut,
  Friedrich-Schiller-Universit\"at Jena, D-07743 Jena, Germany}
 \author{Bertram Klein}\affiliation{Physik Department,
  Technische Universit\"at M\"unchen, James-Franck-Stra{\ss}e 1, D-85748
  Garching, Germany } 
\author{Bernd-Jochen Schaefer}\affiliation{Institut f\"ur Physik, Karl-Franzens-Universit\"at {Graz}, A-8010
  Graz, Austria}\affiliation{Institut f\"ur Theoretische Physik, Justus-Liebig-Universit\"at Gie{\ss}en, D-35392 Gie{\ss}en, Germany}
 
\begin{abstract}
  We investigate the effect of a finite volume on the critical
  behavior of the theory of the strong interaction (QCD) by means of a
  quark-meson model {for $N_\mathrm{f}=2$ quark flavors}.  In
  particular, we analyze the effect of a finite volume on the location
  of the critical point {in the phase diagram}
  existing in our model. In our analysis, we take into account the
  effect of long-range fluctuations with the aid of renormalization
  group techniques. We
  find that these {quantum and thermal} fluctuations,
  absent in mean-field studies, play an import role for the dynamics
  in a finite volume. We show that the critical point is shifted
  towards smaller temperatures and larger values of the quark chemical
  potential if the volume size is decreased. {This behavior persists 
  for antiperiodic as well as periodic} boundary conditions for the quark 
  fields as used in {many} lattice QCD simulations.
\end{abstract}


\maketitle

\section{Introduction}
The detection of the critical endpoint in the phase diagram at finite
temperature and quark chemical potential of the theory of the strong
interaction (QCD) is an inherently difficult problem. In fact, it may
even be the case that {there is no first-order chiral phase
  transition at large chemical potential~\cite{Klimt:1990ws,
    Kitazawa:2002bc, Sasaki:2006ws, Fukushima:2008is,
    Fukushima:2008wg, Schaefer:2008hk, Bratovic:2012qs} and} the
critical endpoint (CEP) of the chiral phase boundary does not exist at
all. The complications arising in the search of the critical point are
manifold.  For example, one of the most important fully
nonperturbative theoretical tools for the exploration of the QCD phase
diagram, namely lattice QCD simulations, suffer from the so-called
sign problem when applied to finite quark chemical potential. Although
powerful techniques have been developed to circumvent this problem
{for smaller chemical
  potentials}~\cite{Fodor:2001au, Fodor:2001pe,
  Fodor:2002km,Allton:2002zi, Allton:2003vx,%
  Allton:2005gk, Gavai:2003mf, Gavai:2004sd,
  Gavai:2008zr,deForcrand:2002ci,
  deForcrand:2002yi,deForcrand:2003bz}, ranging from re-weighting
techniques over Taylor expansions to imaginary-valued quark chemical
potentials, these studies are still restricted to a finite simulation
volume, see, e.g., Refs.~\cite{Philipsen:2005mj, Schmidt:2006us,
  Philipsen:2008gf,deForcrand:2010ys} for reviews.

The effects of a finite volume on the curvature of the chiral phase
transition line at small chemical potentials can be
{determined} from lattice QCD results~\cite{deForcrand:2002ci,
  deForcrand:2002yi,deForcrand:2006pv,Karsch:2003va,Kaczmarek:2011zz,Endrodi:2011gv}.
Recently, the effects of a finite volume~$V=L^3$ as well as of
long-range {fluctuations have been analyzed by means of}
the quark-meson model which serves as a low-energy QCD
model~\cite{Klein:2010tk, Braun:2011iz}. {It was
  found that depending on the pion mass the curvature becomes
  continuously} smaller for decreasing volume sizes and periodic
boundary conditions for the quark fields in the spatial
directions. Decreasing the length of the box below~$m_{\pi}L \lesssim
2$, the curvature then tends to increase again and even exceeds its
infinite-volume value. This very specific behavior can be traced back
to the implementation of periodic boundary conditions. For
antiperiodic boundary conditions, the quarks do not have a spatial
zero mode {and}
the curvature is a monotonically decreasing function of the volume
size.  These results for the curvature are in accordance with the
behavior of low-energy observables, such as the quark condensate, as a
function of the volume size at zero
temperature~\cite{Braun:2005gy,Luecker:2009bs}.  In any case, the
volume dependence of the curvature suggests also that the (chiral) CEP
may be shifted in a finite volume. From the above discussion, for
instance, {it appears most likely} that the critical point is shifted towards
larger values of the chemical potential and smaller temperatures in
case of periodic boundary conditions and~$2\lesssim m_{\pi}L < \infty$. 
However, it may also very well completely disappear in the
small-volume limit.
{The discussion of the volume dependence of the location of the CEP
  also raises a conceptional issue, namely how to define a CEP}
associated with a diverging susceptibility in a finite system. We
shall discuss this in more detail below.

In the present work, we shall focus on the behavior of the QCD
critical endpoint as a function of the volume size{,} with an
emphasis on quark fields with periodic boundary conditions in spatial
directions. {We} expect that this setup is {the most useful one}
to help {in further guiding} present and future lattice simulations
of the QCD phase diagram. For our study, we employ the quark-meson
model.  This model has already proven to be useful in our previous
studies of chiral dynamics at finite temperature 
{or} quark chemical potential in a finite
volume~\cite{Braun:2005fj, Braun:2010vd, Klein:2010tk,
  Braun:2011iz}. {We expect that such a model setup is very useful to}
provide a framework for understanding better the mechanisms of chiral
symmetry breaking in finite systems.  The volume dependence of the
chiral CEP has been studied before with the aid of quark-meson-type
{models in mean-field approximations}
\cite{Palhares:2009tf, Palhares:2011jf, Fraga:2011hi, Palhares:2011zz,
  Bhattacharyya:2012rp}. In the present work, we go beyond the
mean-field {approximation} and include the fluctuations of the
meson fields. {We have found that the fluctuations of bosonic
  fields, in particular fluctuations of the order parameter, are of
  utmost importance for the physics of such models in a finite
  volume~\cite{Braun:2004yk,Braun:2010vd,Klein:2010tk,Braun:2011iz},
  in addition to the effects of fermionic fluctuations.} 
  For a continuous symmetry such as the chiral flavor symmetry, fluctuations
of Goldstone modes restore the symmetry in a finite-volume system in
the absence of explicit symmetry breaking. {This implies} that
{the} phase with spontaneously broken chiral symmetry {does not
  exist at all} in the chiral limit~\cite{Gasser:1987ah}. 
Therefore it is reasonable to expect that the chiral
  phase transition temperature decreases when the current quark mass (pion mass) is decreased for a given fixed 
  volume size. This has also been found in a previous renormalization group study including fluctuation effects
  associated with the Goldstone modes, see Ref.~\cite{Braun:2005fj}.
  Note that a mean-field study cannot capture this effect correctly due to the absence of 
  the Goldstone-mode fluctuations. In fact, it is possible to obtain a finite chiral phase transition temperature
  in mean-field theory in a finite-volume system, even in the chiral limit.
  On the other hand, if our present study of the shift of the CEP in a finite volume is in accordance with
  previous mean-field studies in a given parameter range defined by the dimensionless quantity~$m_{\pi}L$, then
  our results can be considered as a hint that fluctuation effects are at least subdominant in this $m_{\pi}L$-regime for some
  observables.

In {the next} Sec.~\ref{sec:RGsetup} we briefly introduce {the}
renormalization group (RG) setup which underlies our study. The
results for the location of the critical point as a function of the
volume size are then discussed in Sec.~\ref{sec:results}. Finally, our
conclusions and a brief outlook are given in
Sec.~\ref{sec:conclusions}.

\section{Renormalization Group Setup}\label{sec:RGsetup}

For our investigation of critical behavior in a finite volume, we
employ the quark-meson model {for $N_\mathrm{f}=2$ flavors}. 
In our present work, we shall use the following ansatz/truncation
for the RG scale-dependent effective action:
\begin{eqnarray} 
  \Gamma_{k}[\bar q,q,\phi]&=& \int d^{4}x \Big\{
  \bar{q} \left({\rm i}{\partial}\!\!\!\slash + 
    {\rm i}g(\sigma+i\vec{\tau}\cdot\vec{\pi}\gamma_{5}) + {\rm i} \gamma_0 \mu\right)q\nonumber \\ 
  && \qquad\qquad
  +\frac{1}{2} (\partial_{\mu}\phi)^{2}+U_{k}(\phi^2) - c \sigma \Big\}\,,
\label{eq:QM}
\end{eqnarray}
where $\phi^{\mathrm{{T}}}=(\sigma,\vec{\pi})$ {has
  $N_\mathrm{f}^2=4$ components and $q$ represents $N_\mathrm{f}=2$
  quark flavors}. The scale~$k$ denotes the RG scale which is introduced below.
The initial condition for our RG flow study of this effective action is given by the 
so-called classical or ``microscopic" action defined at the scale~$k=\Lambda$.
 This scale is determined by the validity of a hadronic representation of QCD.
The mesonic potential at the UV scale is parameterized by two
couplings, $m^2_\Lambda$ and $\lambda_\Lambda$,
\begin{equation}
  \label{eq:pot_UV} 
  U_\Lambda(\phi^{2}) =
  \frac{1}{2}m_\Lambda^{2}\phi^{2} +
  \frac{1}{4}\lambda_\Lambda(\phi^{2})^{2}
  \,.
\end{equation}
The current quark masses are determined by the term linear in the
field~$\sigma \sim \bar{q}q$ which also renders the pion
fields~$\vec{\pi}$ massive in the low-energy limit.  The bosonic
couplings~{$m_\Lambda$} and~{$\lambda_\Lambda$}, the Yukawa
coupling~$g$, and the $O(4)$-symmetry breaking term~$c$ are parameters
of our model which are used to fit a given set of low-energy
observables, see {the} discussion below. In our model, the order
parameter for chiral symmetry breaking is given by the (vacuum)
expectation value of the {scalar field $\langle
  \sigma\rangle$} which can be identified with the pion decay
constant~$f_{\pi}$. Note that we do not take into account a possible running
of the Yukawa coupling~$g$, as done in, e.g., Refs.~\cite{Jungnickel:1995fp,Berges:1997eu,Berges:2000ew,Braun:2009si,Braun:2011fw,Braun:2012zq,Pawlowski:2014zaa}.
The analysis of these studies suggest that it is justified to neglect the running of this coupling in our study, 
at least for our first RG study of the phase diagram in a finite volume. For more quantitative predictions of the shift of the CEP, of course, this running has to be taken into account eventually.

In the following, we study the RG flow of the order-parameter potential~$U$
{in} leading order of the derivative expansion.
We do not take into account corrections arising from a nontrivial RG
flow of the wave-function renormalizations of the quark and meson
fields.\footnote{To be specific, we consider the wave-function renormalizations $Z_{\phi}$ and $Z_{\psi}$ of the mesons and the quarks to be constant, respectively:
$Z_{\psi}=Z_{\phi}=1$. This implies that the anomalous dimensions associated with these fields are zero, $\eta_{\psi}=\eta_{\phi}=0$.}
{Since we are not aiming
at a quantitatively accurate determination of either the absolute location of the QCD critical point or of the critical behavior as measured by
critical exponents, we consider this to be a justified approximation.} 
However, we would like to emphasize that our present approximation
already includes effects beyond the mean-field limit. In fact, it has
been found that {the critical exponents it yields at the thermal
  phase transition in the quark-meson model already agree} with the
exact values on the 2\% level,\footnote{Note that a detailed error analysis is difficult and is mostly
done by varying and/or optimizing the regularization scheme, by taking into account higher orders in the derivative 
expansion, and by taking into account the full momentum dependence of a given set of $n$-point 
functions, see, e.g., Refs.~\cite{Litim:2002cf,Canet:2002gs,Canet:2003qd,Benitez:2009xg,Litim:2010tt}.} 
see, e.g., Refs.~\cite{Schaefer:1999em,
  Stokic:2009uv, Braun:2010vd,Berges:1997eu,Berges:2000ew,
  Bohr:2000gp}. This observation can be traced back to the {small}
anomalous dimensions associated with these wave-function
renormalizations {found in the} model, see, e.g.,
Refs.~\cite{Berges:1997eu,Berges:2000ew, Bohr:2000gp}. A more
detailed discussion of the relation of the present approximation to
the mean-field approximation in terms of a derivative expansion and a
so-called large-$\Nc$ expansion of the quantum effective
action~$\Gamma$ can be found in Refs.~\cite{Berges:2000ew,
  Braun:2008pi, Braun:2009si, Braun:2011pp, Braun:2012zq}.

Before we turn to the discussion of our {actual} results for the
CEP, we would like to briefly introduce a few more details on our
approach.  In the following we employ the \mbox{Wetterich} equation
{for the derivation of the RG flow equation}
for the order-parameter potential~$U$ \cite{Wetterich:1992yh}:
\begin{equation} 
  \label{eq:Flow_EffAction} 
  \partial_t \Gamma_{k} = \frac{1}{2} \Tr\left\{  \left[\Gamma _{k}^{(2)} + R_k\right]^{-1} \partial_t
      R_k \right\} \ .
\end{equation}
This flow equation describes the change of the quantum effective
action~$\Gamma$ under variation of the RG scale~$k$, {or the RG
`time' $t=\ln (k/\Lambda)$}, and therefore
allows us to interpolate between the {initial}
action~$S\simeq \Gamma_{\Lambda}$ at the UV scale~$\Lambda$ and the
quantum effective action~$\Gamma\equiv\Gamma_{k\to 0}$ in the
limit~$k\to 0$.  To regularize the theory, a regulator function~$R_k$
is included.  For our finite-temperature and finite-volume studies, it
is convenient to choose a dimensionally reduced regulator function,
i.e. a so-called spatial regulator which only regularizes the spatial
momentum modes~\cite{Litim:2006ag, Blaizot:2006rj}. Our particular
choice for the regulator function is a close relative of the optimized
regulator function~\cite{Litim:2000ci,
  Litim:2001up,Litim:2001fd,Pawlowski:2005xe}.
  We would like to add that the use of the dimensionally reduced regulator 
  at finite temperature does not cause any conceptional problem in the local potential approximation
  underlying the present work. At next-to-leading order in the derivative expansion,
  complications arise from the fact that this class of regulators breaks explicitly Poincare invariance. 
  However, this can then be handled by suitably adjusting the initial conditions of the RG flow~\cite{Braun:2009si}.

{Inserting} the {truncation} \eqref{eq:QM} into
Eq.~\eqref{eq:Flow_EffAction}, we obtain the flow equation for the
RG-scale dependent order-parameter potential~$U_k$ in a finite
cubic volume~$V=L^3$, see Refs.~\cite{Braun:2010vd,Braun:2011iz}:
  \begin{eqnarray}
    \label{eq:fv_ft_flow_equation}
   \partial_t U_k(\phi^2)
    &=& k^5 \Bigg[ \frac{3}{E_\pi} 
    \left( \frac{1}{2}+n_B(E_\pi) \right){\mathcal B}_{\text{p}}(kL) \nn\\
    && \qquad +\frac{1}{E_\sigma} \left(
      \frac{1}{2}+n_B(E_\sigma)\right){\mathcal
      B}_{\text{p}}(kL) \nn\\
   && \qquad\quad -\frac{2 N_c N_f}{E_q} \Big( 1-n_F(E_q,\mu) \nn\\ 
   && \qquad\quad\quad - n_F(E_q,-\mu)
    \Big){\mathcal B}_{\ell}(kL) \Bigg] \,,
  \end{eqnarray}
where the effective quasi-particles energies of the mesons and quarks are {defined} as
\begin{equation}
E_i=\sqrt{k^2+M_i^2} \; , \quad i \in \{ \pi, \sigma, q \} \,.
\end{equation}
The effective squared meson masses are determined by derivatives of the potential~$U_k$,
\begin{eqnarray} 
  M_{\pi}^2 = 2 \frac{\partial
    U_k}{\partial \phi^2}\,,&&\quad
  M_{\sigma}^2 = 2 \frac{\partial U_k}{\partial \phi^2} + 4 \phi^2
  \frac{\partial^2 U_k}{\partial (\phi^2)^2}\,,\label{eq:m2}
\end{eqnarray} 
whereas the quark mass is directly related to the Yukawa coupling $g$,
\begin{equation} 
  M_{q}^2 = g^2 \phi^2.
\end{equation}
{Note that, strictly speaking, the
  quantities~$M_{\pi}$, $M_{\sigma}$, and~$M_{q}$ are the physical
  masses of the associated particles {only} if evaluated {in} the ground
  state of} the theory.  The thermal dynamics of the theory is
controlled by the bosonic and fermionic occupation numbers,
\begin{equation}
  n_B(E)=\frac{1}{e^{E/T}-1} \; , \quad n_F(E,\mu)=\frac{1}{e^{
      (E-\mu)/T}+1} \,,
\end{equation}
whereas the volume dependence is governed by the
mode counting functions ${\mathcal B}_{\ell}$:
\begin{equation} 
  {\mathcal B}_{\ell}(kL)=\frac{1}{(kL)^3} \sum_{\vec{n} \in \mathbb{Z}^3} 
  \theta\!\left( (kL)^2    
    - (2n+\delta_{\text{{ap}},\ell})^{\,2}\pi^2\right)\,.  
\end{equation}
Here, $\vec n$ labels a three-dimensional vector of integers and the
label~$\ell$ refers to antiperiodic ({ap}) and periodic (p)
boundary conditions, respectively. Note that, due to our choice for
the regulator function, the spatial and thermal contributions in the
flow equation factorize, which facilitates the numerical
{evaluation~\cite{Litim:2006ag}.}

From a phenomenological point of view, the mode counting functions
already reflect the fact that the dynamics of the theory in the
small-volume limit is only governed by the spatial zero mode in the
case of periodic boundary conditions. For antiperiodic boundary
conditions, on the other hand, we observe that these functions tend to
zero for~$kL \to 0$. Thus, the fields effectively become static and
condensation of quark-antiquark pairs is suppressed. In the infinite-volume
limit ($L\to \infty$), we find that these functions approach a finite
number which {depends only} on the chosen regularization
scheme. {We} recover the known flow equation for the chiral
order-parameter potential of the quark-meson model {in this}
limit~\cite{Braun:2003ii,Schaefer:2004en}.

The flow equation~\eqref{eq:fv_ft_flow_equation} for the order-parameter potential is a partial differential equation in the
RG scale~$k$, where~$0\leq k \leq \Lambda$, and the field variable~$\phi$.
In the present study, we solve it by discretizing the field variable~$\phi$, see, e.g., Ref.~\cite{Schaefer:2004en} for details. 
For a review on results from such a direct numerical solution of this type of flow equation for the potential
and, in particular, for results concerning the phase diagram of QCD low-energy models in the infinite-volume limit,
we refer the reader to Ref.~\cite{Schaefer:2006sr}. 

For our numerical study of the phase diagram, we need to fix the parameters of our model, namely~$m_{\Lambda}$, $\lambda_{\Lambda}$, the Yukawa
coupling~$g$, the UV scale~$\Lambda$, and the symmetry breaking parameter~$c$. 
As already mentioned above, we use these parameters to fit a given set of low-energy observables,
namely the pion mass~$m_{\pi}$, the constituent quark mass~$m_{q}$, and the pion decay constant~$f_{\pi}$. 
Clearly, the set of 
parameters {determined in this way} is by no means unique.
Loosely speaking, this parameter ambiguity leaves its trace in the phase structure of the quark-meson model. 
{As} has been pointed out in Ref.~\cite{Schaefer:2008hk}, the chiral CEP can be shifted almost arbitrarily by varying the mass of the 
sigma meson. Here, we therefore pursue the following strategy: we choose a set of initial conditions such that we find
a chiral CEP at a certain position in the infinite-volume phase diagram. 
{Since} we are not aiming at a quantitative determination of the position of the CEP, the absolute values of its coordinates play only a {secondary} role.
Once the parameters have been fixed,
we vary the box size and follow the
shift of the CEP. We have checked for various sets of parameters that the qualitative behavior of the CEP as a function of the 
volume size indeed appears to be `universal'. {To be specific, in the calculation presented here,} 
 we have used~$\Lambda=1\,\text{GeV}$, $m_{\Lambda}=0.881\,\text{GeV}$, $\lambda_{\Lambda}=0$,
and~$c=5\cdot 10^{-4}\,\text{GeV}^3$ {for the model parameters}. At zero temperature, this yields~$m_{q}=318\,\text{MeV}$, $m_{\sigma}=461\,\text{MeV}$, and~$m_{\pi}=75\,\text{MeV}$.
From a phenomenological point of view, {the} pion mass is clearly too small. Nevertheless, we have chosen this value since
it currently appears to be the smallest accessible value for the pion mass in lattice simulations~\cite{Ejiri:2009ac} and we expect that the 
effect of the long-range fluctuations in a finite volume is most pronounced for small pion masses. {While the results for a pion mass of $m_\pi \approx 140$ MeV are similar, this choice leads to a qualitatively clearer picture of the finite-volume effects.\footnote{
 Note that, if we tune the parameters such that the pion mass eventually approaches its phenomenological value,~$m_{\pi}\approx 140\,\text{MeV}$ 
 for vanishing temperature and chemical potential, 
 we find that the CEP in infinite volume is already shifted to very small values of the temperatures. For finite volumes, we find that the 
 CEP is then shifted in the same way as for smaller pion masses,
 see our discussion below. However, the fact that this takes place at very low temperatures makes a numerical study presently very cumbersome.}}

\section{Critical Endpoint and Finite-Volume Effects}\label{sec:results}

Let us {now} discuss effects of a finite volume on the location of
the chiral CEP. {The CEP} is associated with a residual $Z(2)$
symmetry. {For finite pion masses, {it} is uniquely defined in the
infinite-volume limit by a diverging (longitudinal)
susceptibility~$\chi_{\sigma}\sim \partial \sigma_0/\partial c$, where $\sigma_0$ 
is the order parameter for chiral symmetry breaking\footnote{Note that $\sigma_0$ can be identified with
the pion decay constant in our quark-meson model. Moreover, we have $\sigma_0\sim \langle \bar{\psi}\psi\rangle$.}
and $c$ is a measure for the explicit symmetry breaking, see Eq.~\eqref{eq:QM}. 
In our model study, $\chi_{\sigma}$ assumes a simple form, 
see, e.g., Refs.~\cite{Braun:2010vd,Schaefer:2006ds}:
\be
\chi_{\sigma} = \frac{1}{M_{\sigma}^2}\,.
\ee
where $M_{\sigma}^2$ is defined in Eq.~\eqref{eq:m2}.}
The corresponding transversal susceptibility associated with the
Goldstone modes, i.e. pion modes, does not provide us with new
information for {the} present study, {since} it is directly
related to the chiral order parameter.
\begin{figure*}
\includegraphics[clip=true,width=\columnwidth]{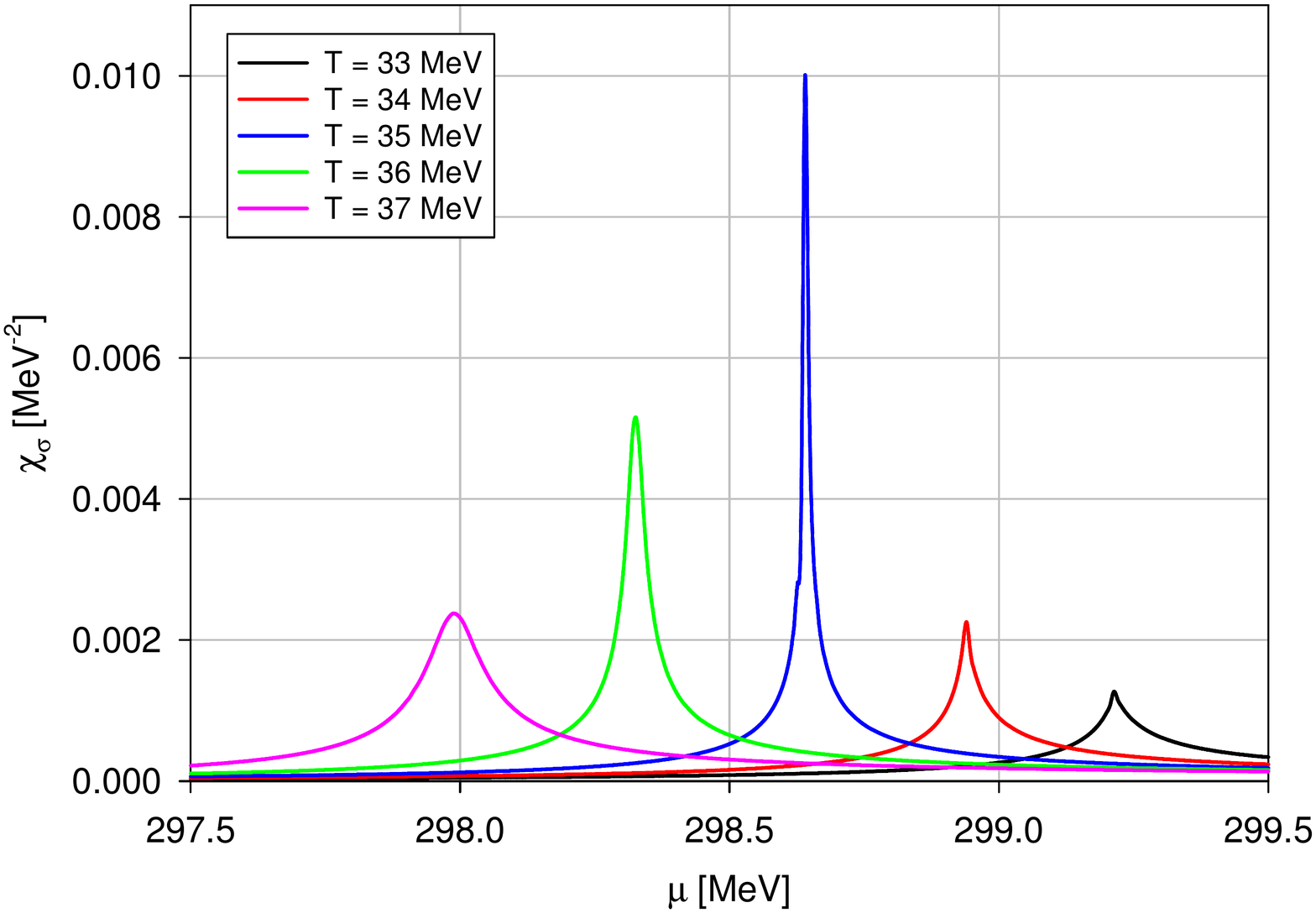}
\includegraphics[clip=true,width=\columnwidth]{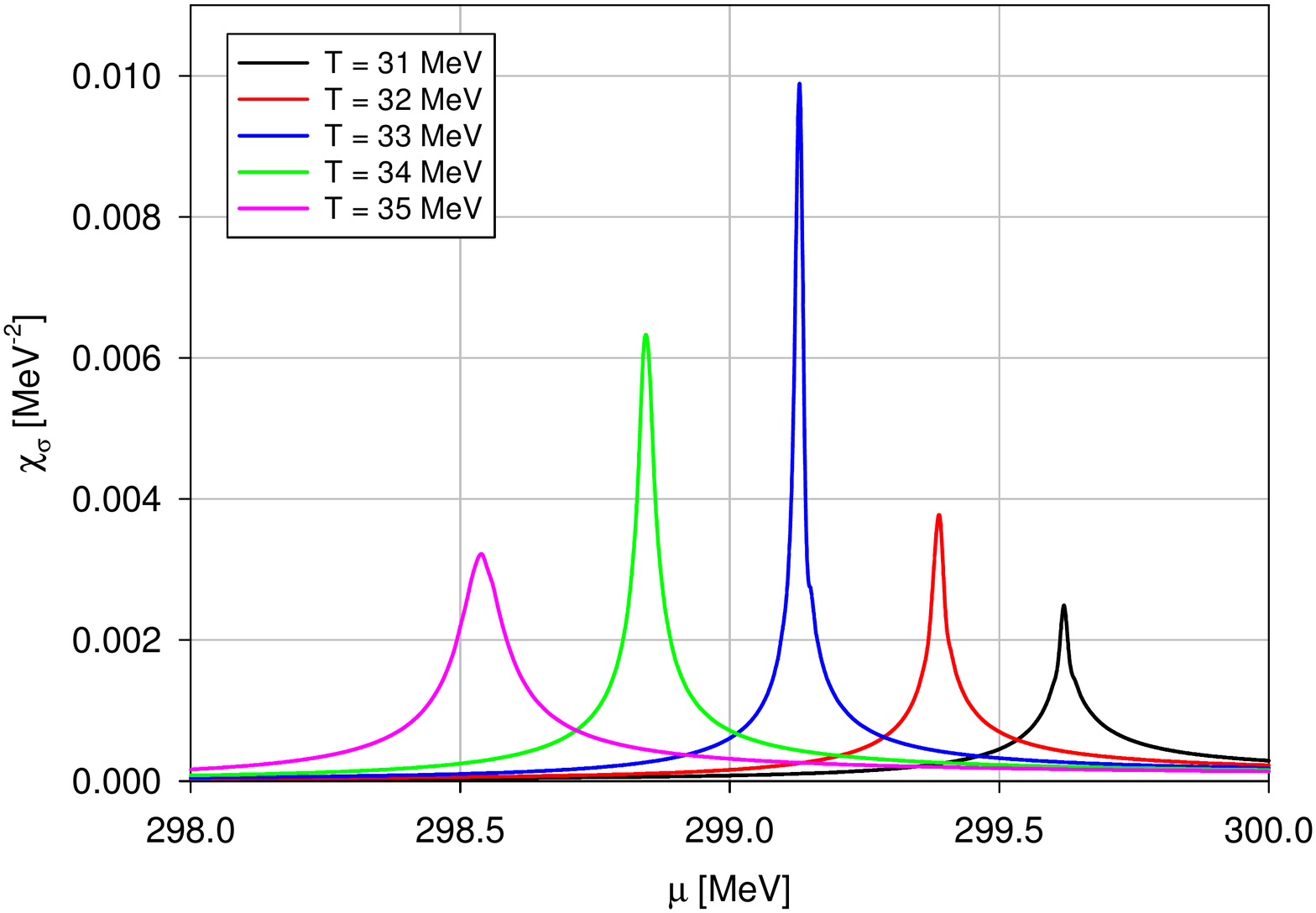}
\includegraphics[clip=true,width=\columnwidth]{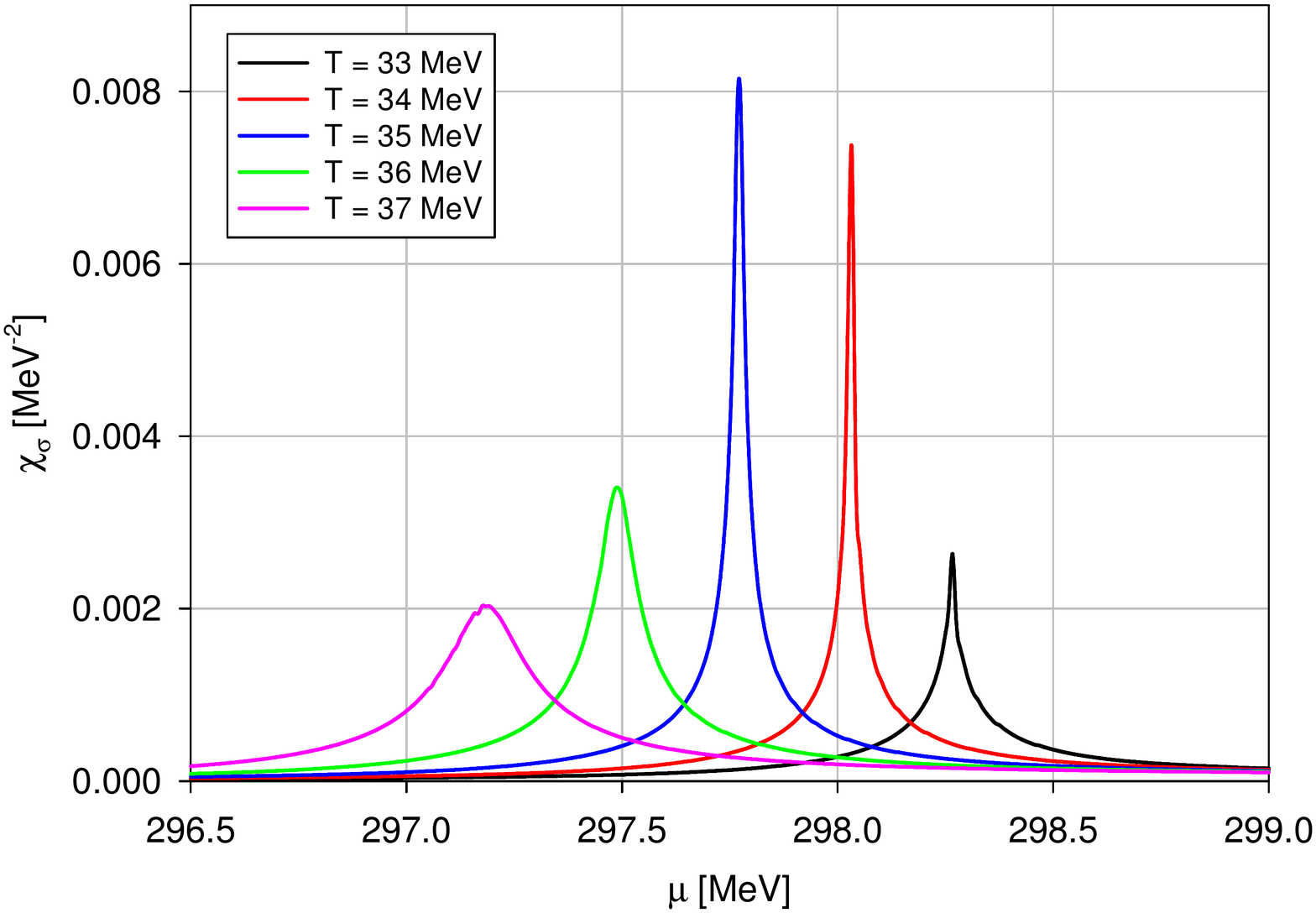}
\includegraphics[clip=true,width=\columnwidth]{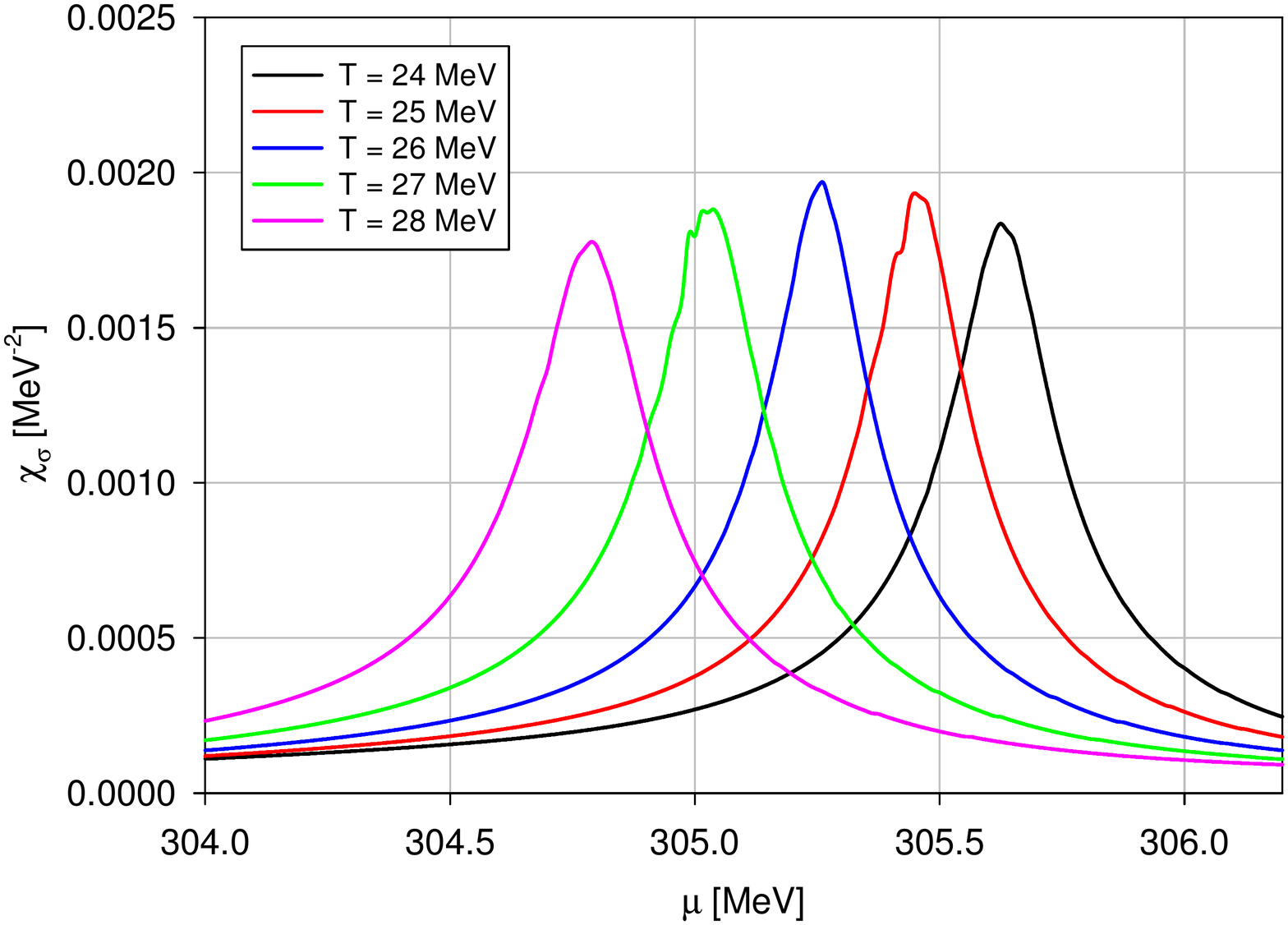}
\includegraphics[clip=true,width=\columnwidth]{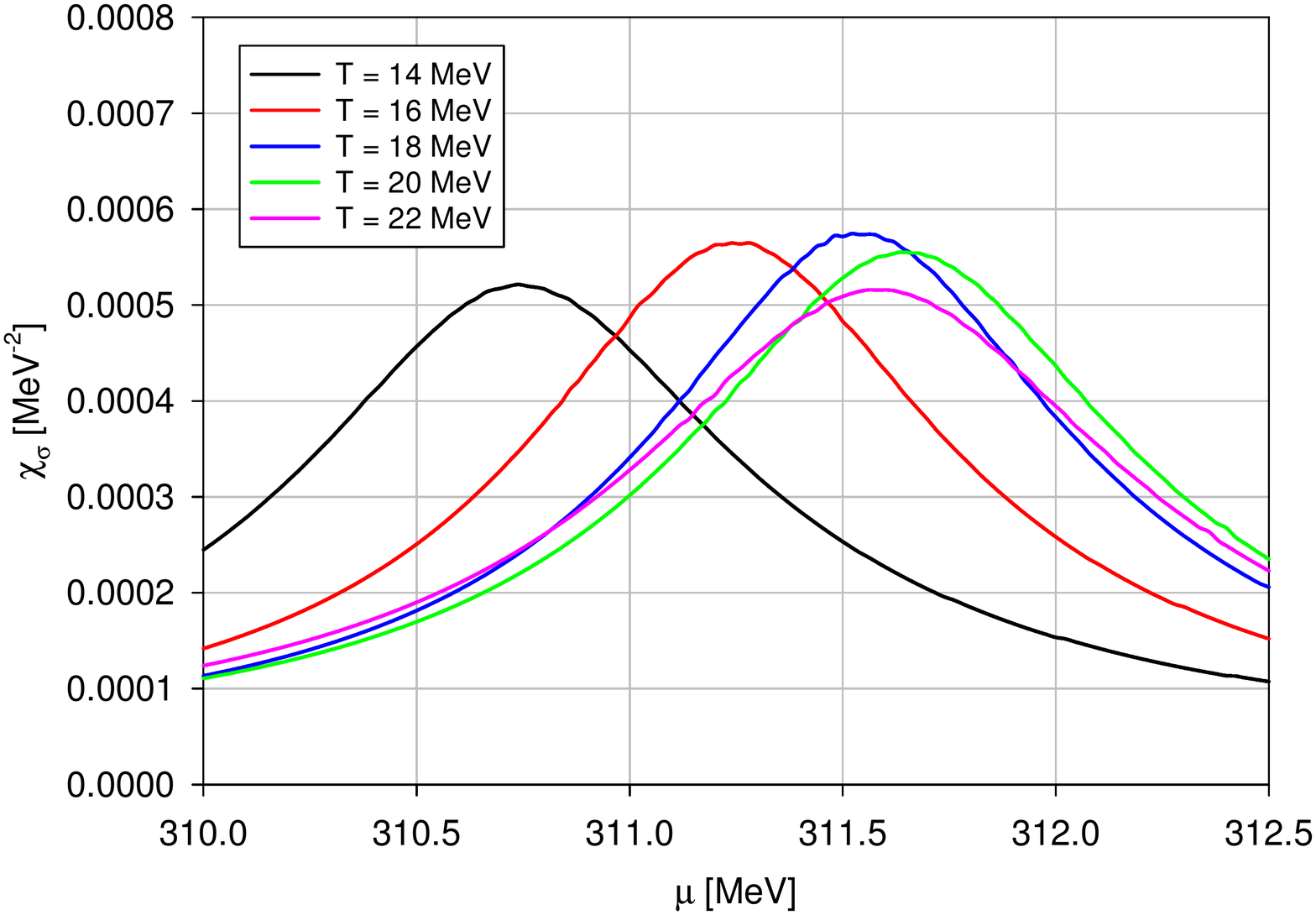}
\includegraphics[clip=true,width=\columnwidth]{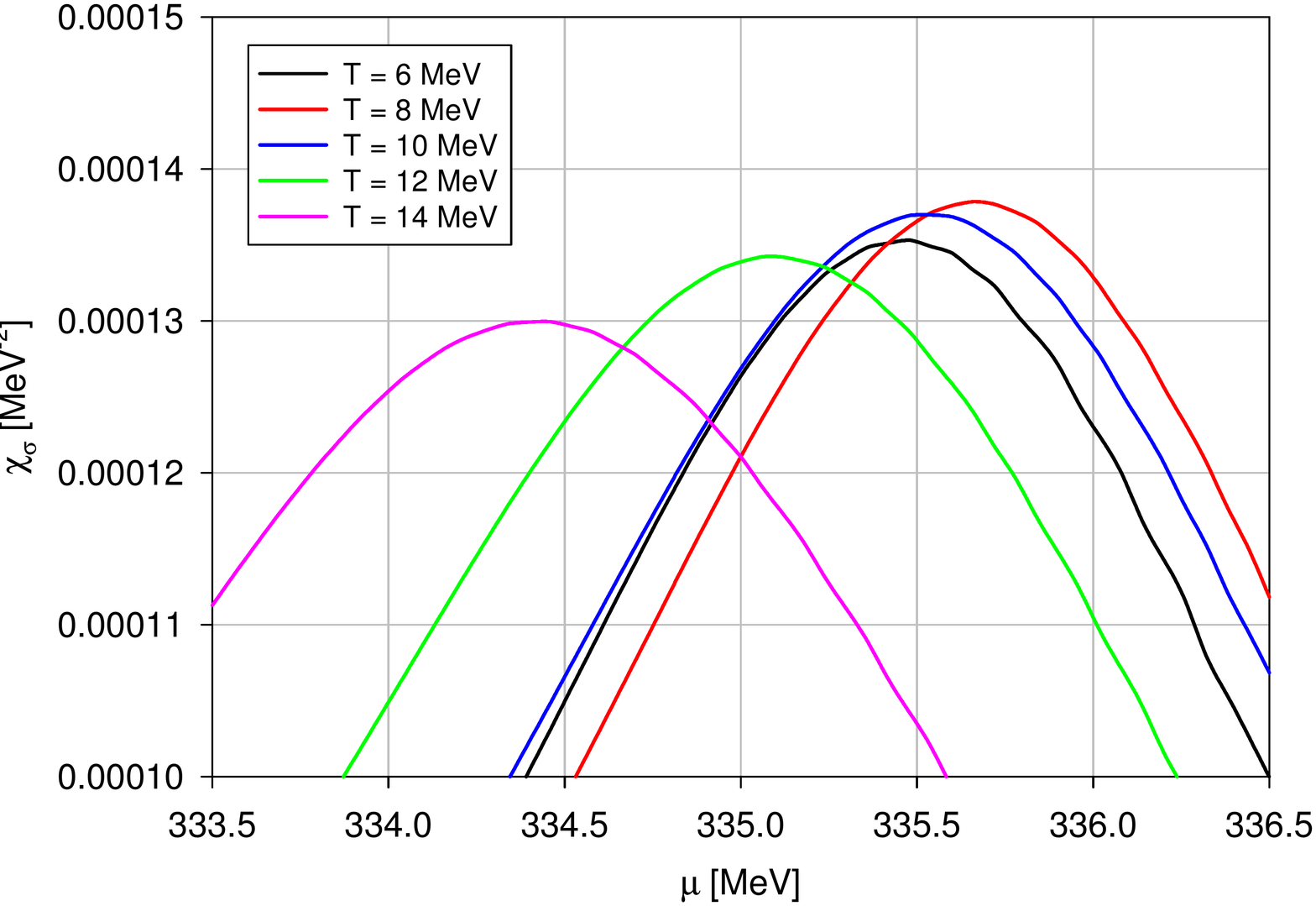}
\caption{(color online) Chiral {(longitudinal)}
  susceptibilities~$\chi_{\sigma}=\frac{1}{M_{\sigma}^2}$ as a
  function of the quark chemical potential~$\mu$ for various
  temperatures and volume sizes ranging from~$L\to\infty$ from the top
  left panel to $L=4\,\text{fm}$ in the bottom right panel as obtained for 
  periodic boundary conditions for the quark fields in spatial directions.}
\label{fig:susz} 
\begin{picture}(0,0)(80,40)
\put(3,585){\Large $L\to\infty$}
\put(245,585){\Large $L=10\,\text{fm}$}
\put(-3,410){\Large $L=8\,\text{fm}$}
\put(245,410){\Large $L=6\,\text{fm}$}
\put(-3,235){\Large $L=5\,\text{fm}$}
\put(245,235){\Large $L=4\,\text{fm}$}
\end{picture}
\end{figure*}

With the parameter set given in the previous section, we find a
diverging (longitudinal) susceptibility associated with a chiral CEP
at
\be
{\big(\mu_{\rm CEP}^{(\infty)},T_{\rm CEP}^{(\infty)}\big)\approx (298.6\pm 1\,\text{MeV},35.0\pm 1\,\text{MeV})\,,}
\ee
see also top left panel of Fig.~\ref{fig:susz}. {The errors arise from an 
uncertainty in our numerical determination of the coordinates of the CEP and are identical for both our infinite volume study
and our finite-volume studies. These errors do not account for systematic errors associated with, e.g., our truncation
for the effective action. In particular,
we would like to}
stress again that the results for the location of the CEP in
quark-meson- or Nambu--Jona-Lasinio-type models in general suffer from
a parameter ambiguity~\cite{Schaefer:2008hk}. However, this ambiguity
{affects the} present study {very little, since} we are only
interested in the shift of the position of the CEP in a finite volume
relative to its infinite-volume coordinates.

As we have already indicated above, the fluctuations of the meson
fields play a prominent role in studies of the chiral critical
behavior in a finite volume.
{This is a consequence of the appearance of the pions as the light
  Goldstone modes in the breakdown of the chiral symmetry.}  In
absence of an explicit symmetry breaking term, these fluctuations
restore the chiral symmetry in a finite volume in the long-range
{(infrared)} limit~\cite{Gasser:1987ah}.  A further
complication related to {the} finite volume {in a study of
  critical behavior} is to find a proper definition of a CEP {which
  is in the infinite-volume limit} associated with a diverging
susceptibility. {In a finite volume,} the susceptibilities are
expected to be bounded from above.  To be more precise, the magnitude
of the susceptibility is assumed to scale with the volume size {like}
\be
\chi_{\sigma} \sim L^2 \sim V^{\frac{2}{3}}\,.
\ee
This already follows from a simple dimensional analysis.{\footnote{{The relation between the maximal value of the susceptibility
and the box size {$\chi_\sigma \sim L^{\gamma/\nu}$} may receive corrections, depending on the values of the critical exponents. 
In particular, since $\gamma/\nu = 2-\eta$, simple dimensional analysis gives the correct result only in the absence of an anomalous dimension. 
This holds in our present approximation.}}}
Thus, the susceptibilities are necessarily finite in a finite volume.
In order to study the volume dependence of the position of the CEP, we
therefore first require a definition of the CEP in a finite volume.
For a given finite volume~$V=L^3$, our discussion suggests that the
CEP can be defined
{as the point in the $(\mu, T)$-plane with the
  coordinates~$(\mu_{\rm CEP}^{\rm max.},T_{\rm CEP}^{\rm max.})$
  where the maximum value of the susceptibility~$\chi_{\sigma}^{\rm
    max.}$ is attained:} \be \chi_{\sigma}^{\rm max.}(\mu_{\rm
  CEP}^{\rm max.}(L),T_{\rm CEP}^{\rm max.}(L)) = \max_{(\mu,T)}
\chi_{\sigma}(\mu,T, L)\,.  \ee
In the large-volume limit, we then have
\be
\lim_{L\to\infty}(\mu_{\rm CEP}^{\rm max.}(L),T_{\rm CEP}^{\rm max.}(L))
=\big(\mu_{\rm CEP}^{(\infty)},T_{\rm CEP}^{(\infty)}\big)\,,
\ee
as it should be. In the following we shall use the coordinates~$(\mu_{\rm CEP}^{\rm max.},T_{\rm CEP}^{\rm max.})$ 
to locate the CEP in a finite volume, {with the additional constraint
that the \mbox{so-determined} CEP has to be continuously connected to the infinite-volume
CEP, and thus to the chiral crossover, under a change of the volume size, 
as also shown in Fig.~\ref{fig:pd}. This requirement becomes necessary for volume sizes of
$L\lesssim 5\,\text{fm}$ and periodic boundary conditions where we observe that a second CEP develops in the 
phase diagram at smaller chemical potential and temperatures, with values of the susceptibility
that may exceed those of the CEP at larger chemical potential.\footnote{This second CEP has been studied and
analyzed in, e.g., Ref.~\cite{Schaefer:2004en} but is not at the heart of the present study.}
\begin{figure}
\includegraphics[clip=true,width=\columnwidth]{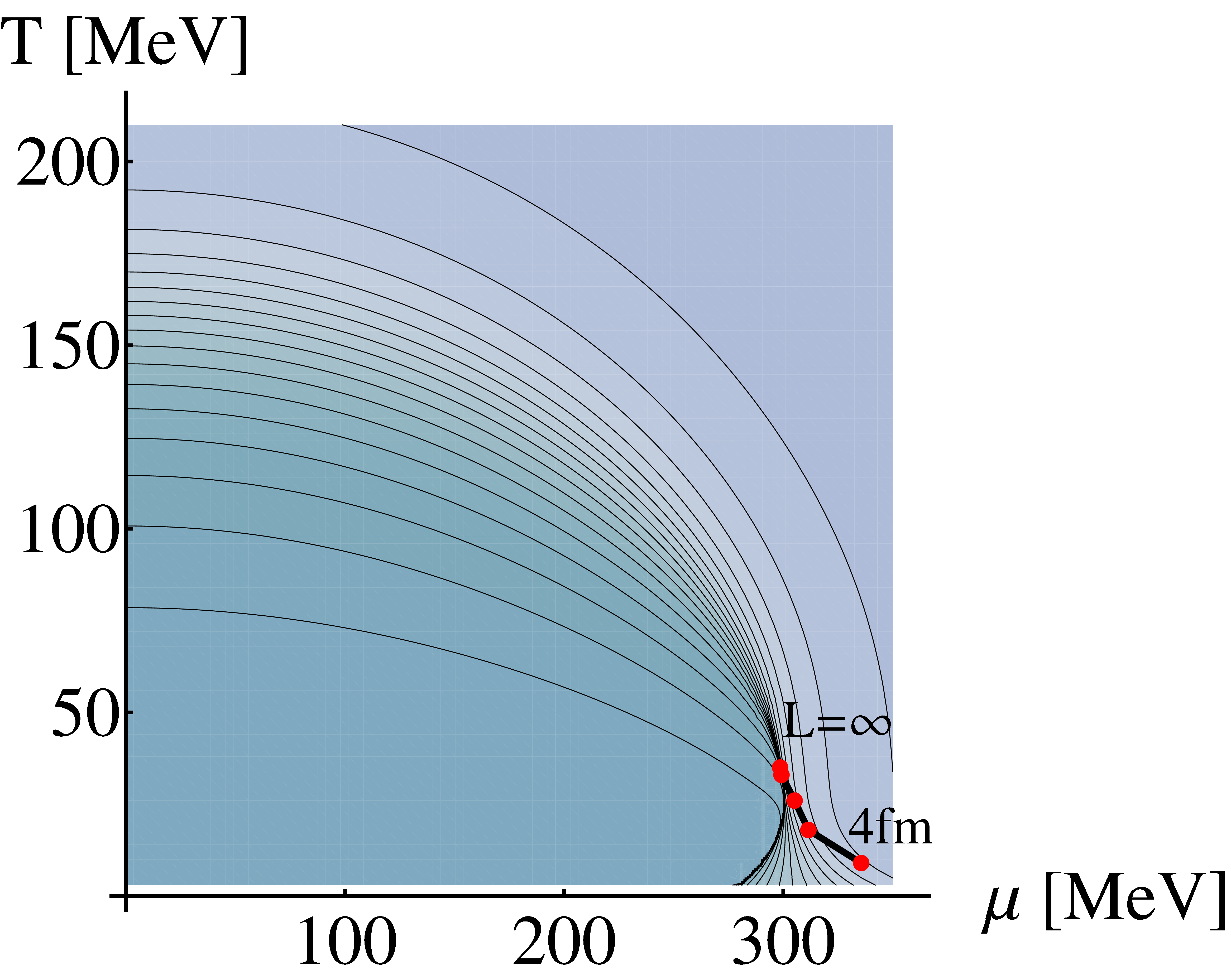}\llap{\makebox[3.5cm][l]{\raisebox{3.5cm}{\includegraphics[height=3cm]{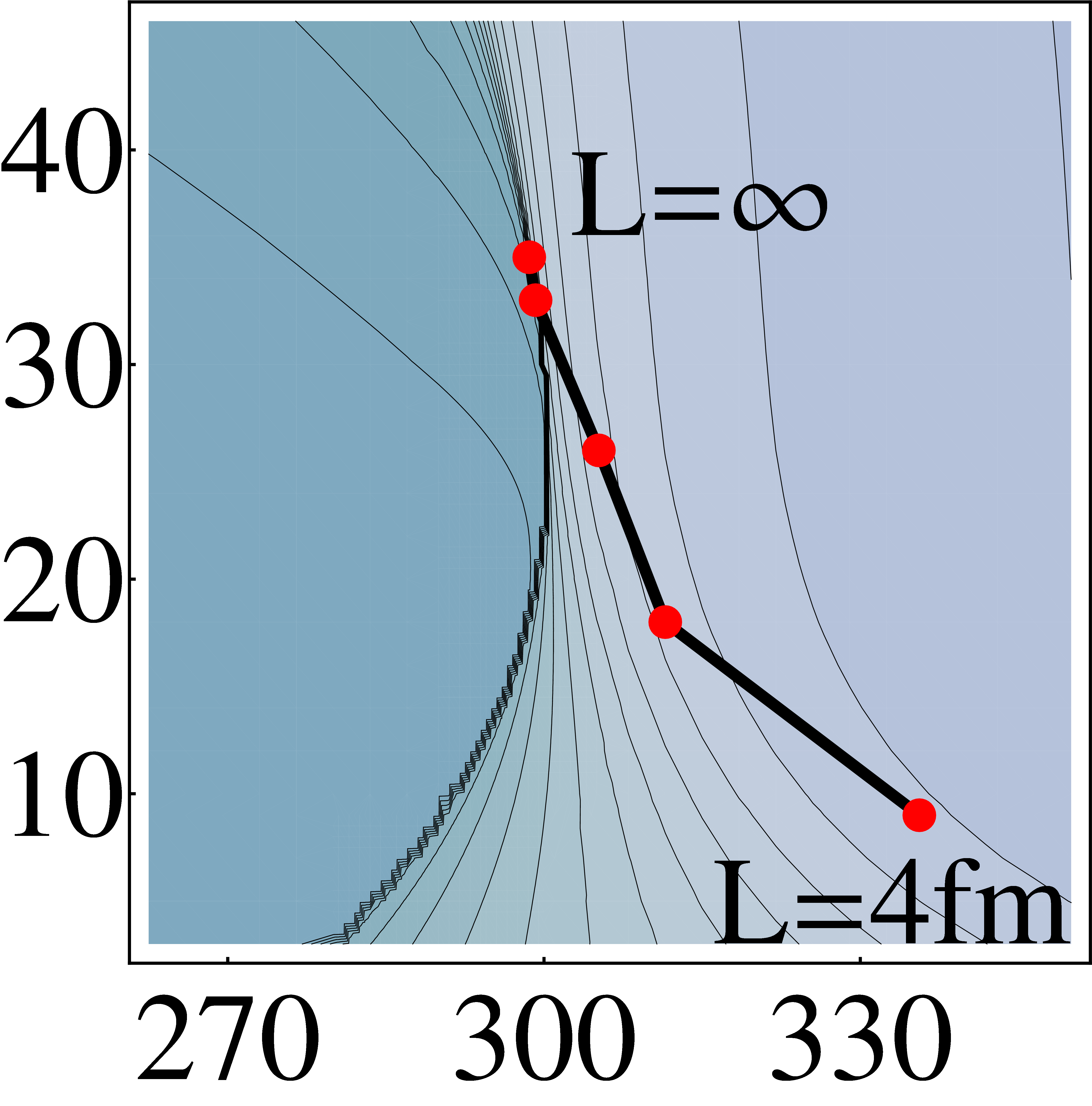}}}}
\caption{(color online) Contour plot of the magnitude of the chiral
  order parameter~$\sigma_0=f_{\pi}$ in the plane spanned by
  temperature and quark chemical potential {for
  $L$ towards infinity}: The order-parameter
  decreases from dark to light shading. The
  inlayed (red) dots show the behavior of the CEP (associated with a
  Z($2$) symmetry) as a function of the volume size for periodic
  boundary conditions for the quark fields in spatial directions. We
  observe that the CEP is shifted towards larger values of the
  chemical potential but smaller values of the temperature
  {for smaller volumes}.}
\label{fig:pd}
\end{figure}

In Fig.~\ref{fig:susz}, we show our results for the longitudinal
susceptibility~$\chi_{\sigma}$ for various temperatures~$T$ and volume
sizes (ranging from~$L\to\infty$ from the top left panel to
$L=4\,\text{fm}$ in the bottom right panel) as a function of the quark
chemical potential~$\mu$. In the infinite-volume limit, we indeed
observe that the susceptibility along the {crossover}
line increases when we increase the quark chemical potential until it
diverges at the CEP. Comparing the panels for the various box sizes,
we also find that the maximum values of the susceptibilities 
{decrease with smaller box size}, in accordance with the
expectations.  {In the first panel of Fig.~\ref{fig:susz}, we show
  the susceptibility for $L\to \infty$ close but not exactly at the
  CEP. The maximum value is limited by $\chi_\sigma^{\text{max.}} \sim
  \frac{1}{k_\mathrm{IR}^2}$ since, for technical reasons, the RG flow
  is evaluated only for $k \to k_{\mathrm{IR}}$ with a small but
  finite value $k_{\mathrm{IR}} \approx 20$~MeV.}  Most importantly,
however, we observe that the position of the CEP is, {within the errors}, shifted towards
smaller values of the temperature and {larger} values of the
chemical potential for decreasing box size.\footnote{
  Note that we have cross-checked our results for the location of the CEP as obtained from an analysis
  of the susceptibility by directly studying the shape 
  of the chiral order-parameter potential.}
To be specific, the CEP
is shifted from~$\big(\mu_{\rm CEP}^{(\infty)},T_{\rm
  CEP}^{(\infty)}\big) {\approx (298.6\pm 1\,\text{MeV},35.0\pm 1\,\text{MeV})}$ for~$L\to\infty$ {to~$\big(\mu_{\rm
  CEP}^{\text{max.}},T_{\rm
  CEP}^{\text{max.}}\big)=(335.6\pm1\,\text{MeV},9.0\pm1\,\text{MeV})$
for~$L=4\,\text{fm}$.} In other words, $\mu_{\rm CEP}^{\text{max.}}$
increases by about $10\%$, whereas~$T_{\rm CEP}^{\text{max.}}$
decreases by about~$75\%$ in the considered range of volume
{sizes, see also Tab.~\ref{tab:CEP} for values of the relative
  shift of the CEP for several other volume sizes.}
Thus, the temperature coordinate of the CEP appears to be {affected more strongly} by the finite volume than 
the one associated with the quark chemical potential.
In Fig.~\ref{fig:pd}, we illustrate {the} results for the shift of the coordinates of the CEP in our model study.
{For $L\lesssim 4\,\text{fm}$, we do not find a CEP any more, at least none compatible with our definition.}
In any case, we have checked that the direction of the finite-volume shift of the CEP does not depend on our choice for the model parameters. The 
latter affects our results only on a quantitative level.
\begin{table}[t!]
\begin{tabular}{|c|c||c|c|}
\hline 
$L\,[\text{fm}]$&
$m_{\pi}L$&
$\Delta\mu_{\rm CEP}$&
$\Delta T_{\rm CEP}$
\tabularnewline
\hline
\hline 
$10$&
$3.75$&
{$0.00(1)$}&
{$-0.06(6)$}
\tabularnewline
\hline 
${8}$&
${3.00}$&
{$0.00(1)$}&
{$\phantom{-}{0.00(6)}$}
\tabularnewline
\hline 
$6$&
$2.25$&
{$0.02(1)$}&
{$-0.26(5)$}
\tabularnewline
\hline 
$5$&
$1.88$&
{$0.04(1)$}&
{$-0.49(4)$}
\tabularnewline
\hline 
$4$&
$1.50$&
{$0.12(1)$}&
{$-0.74(4)$}
\tabularnewline
\hline
\end{tabular}
\caption{{Relative shifts}~$\Delta\mu_{\rm CEP}=(\mu_{\rm CEP}^{\rm max.}/\mu_{\rm CEP}^{(\infty)}-1)$ 
and~$\Delta T_{\rm CEP}=(T_{\rm CEP}^{\rm max.}/T_{\rm
  CEP}^{(\infty)}-1)$ of the {CEP} coordinates{~$(\mu_{\rm CEP}^{\rm max.},T_{\rm CEP}^{\rm max.})$}
in a finite volume~$V=L^3$ {compared to} its position{~$\big(\mu_{\rm CEP}^{(\infty)},T_{\rm CEP}^{(\infty)}\big)$} in the infinite-volume limit. Here, we have chosen
$m_{\pi}=75\,\text{MeV}$ which refers to the
pion mass in the infinite volume limit for~{$(\mu, T)=(0, 0)$}.
For the quantum fields, we have chosen periodic boundary conditions in spatial
directions. {The errors arise from the uncertainty in determining the position of the maximum value of the susceptibility, see main text for details.}
\label{tab:CEP} }
\end{table}

Our results show that the position of the CEP depends strongly on the box size. The fact that
the position of the CEP is shifted towards larger values of the chemical potential and lower values of the temperature
{might be considered helpful} from the point of view of lattice QCD simulations. 
{If the observed shift of the CEP in our model study 
carries over} to lattice QCD simulations,
then it implies that, in a finite volume, the CEP is pushed
into a domain of the phase diagram which is very difficult to access with such simulations. 
 
Our results also indicate that comparatively large volume sizes are
required to {effectively resolve} the position of the CEP in the
infinite-volume limit. To be specific, we find that the position of
the CEP in the infinite-volume limit is only reached for~$L\gtrsim
10\,\text{fm}$. However, {this} statement depends on our choice for
the model parameters which {ultimately} determines the position of
the CEP in the continuum.  {The scale~$L$ is in a broad sense in
  competition with the two other scales~$T$ and~$\mu$ which appear
  also in the quark propagator. All three scales can be considered
 {as} cutoff scales for the low-momentum modes of
  the quark fields.}  If therefore
\be
1/L \ll {M}_{\rm{min}}\,,
\ee
where 
\be
{M}_{\min}=\min\big(\mu_{\rm CEP}^{(\infty)},T_{\rm CEP}^{(\infty)}\big)\,,
\ee
then we expect that the position of the CEP remains effectively unchanged, 
{since $L$ does not impose an additional cutoff}.\footnote{Note that the 
pion mass scale and the scale associated with the pion decay constant have been implicitly taken into account
in this analysis as these quantities are parameters of the model which essentially determine the location of the CEP in the infinite volume
limit.} If, on the other hand, $1/L$ is of the order of~${M}_{\rm min}$,
\be
1/L \sim {\mathcal O}({M}_{\rm min})\,,
\ee
then we expect finite-volume effects on the CEP to become important. Since the actual position of the CEP depends on our
choice for the parameter set, the smallest value of~$L$, for which a significant shift of the position of the CEP is found, is
clearly parameter-dependent. From the point of view of lattice QCD simulations, these considerations imply that the shift of
the CEP due to the presence of a finite volume is only relevant for small box sizes,
if the possibly existing CEP in the QCD phase diagram is located
at smaller values of the chemical potential and larger values of the temperature. In other words,
we expect that {in this case} the infinite-volume position of the CEP {can be reached already} for small box sizes.

Let us finally speculate about phenomenological implications of our results. The
role of finite-size effects in the search for the CEP in heavy-ion collision experiments has been previously discussed
in, e.g., Refs.~\cite{Palhares:2009tf,Palhares:2011jf,Fraga:2011hi,Palhares:2011zz}. Here, we would like to simply comment 
on how the boundary conditions for the quark fields in spatial directions affect our results. In addition to periodic boundary conditions,
we have also studied the shift of the CEP in a finite volume in the presence of antiperiodic boundary conditions for the quarks. We also find in this case
that the CEP is shifted to lower temperatures but larger values of the chemical potential and eventually disappears for~$L\lesssim 4\,\text{fm}$.
Although we expect neither periodic nor antiperiodic boundary conditions to be at work in the expanding fireball in a heavy-ion collision experiment and
although we also do not claim that the observed direction of the shift of the CEP in a finite volume 
holds for general boundary conditions and general geometries of the volume, 
the observed finite-volume shift of the CEP may at least be considered as a possible scenario that could take place in the experiments. Depending
on the experimental setup, traces of the CEP in the experimental data may therefore be found at different coordinates~$(\mu,T)$.

\section{Conclusions and Outlook}\label{sec:conclusions}
Using non-perturbative functional RG techniques, we have computed the shift 
of the CEP of a quark-meson model in a finite volume for periodic as well as antiperiodic
boundary conditions for the quark fields in spatial directions. In our study, we have
also included fluctuations of the meson fields by means of a derivative expansion of the effective action.
The effect of these fluctuations have not been taken into account in recent mean-field 
studies~\cite{Palhares:2009tf,Palhares:2011jf,Fraga:2011hi,Palhares:2011zz,Bhattacharyya:2012rp}.
However, they play an important role in finite-volume {systems}, as they tend to restore the chiral
symmetry in a finite volume {in}  the limit of small quark masses.

The model underlying our studies does not contain gluonic degrees of
freedom and it is not confining. However, it can be considered as an
effective low-energy model for dynamical chiral symmetry breaking
which allows us to analyze the effects of a finite volume on the
chiral dynamics in simple terms. {We} have found that the CEP in a
finite volume is shifted towards smaller temperatures and larger
values of the chemical potential when the volume is decreased, in
accordance with earlier mean-field NJL model
studies~\cite{Palhares:2009tf,Palhares:2011jf,Fraga:2011hi,Palhares:2011zz}.
The volume size below which the CEP is significantly shifted depends
on the actual position of the CEP in the infinite-volume limit. The
same holds for the volume size below which the CEP
disappears. Interestingly, we have found that these qualitative
aspects are present independent of our choice for the {spatial}
boundary conditions (periodic or antiperiodic) for the quark fields.
The fact that the shift of the CEP towards lower temperatures and
{larger} chemical potentials may already set in for comparatively large
volume sizes (depending on the coordinates of the CEP in the
infinite-volume limit) {could be a hint towards a further
  complication in} the search {for} the CEP with lattice QCD
simulations.

{While} our present work focuses on chiral aspects of the QCD phase diagram, 
the inclusion of dynamical gauge degrees of freedom {in this} RG study following Ref.~\cite{Braun:2009gm} (see 
Refs.~\cite{Litim:1998nf,Gies:2006wv} for reviews) {would represent} an interesting extension. This {opens} the possibility
to study the interplay of chiral and confining dynamics in a finite volume. {A considerable 
volume dependence is indeed also expected in the gauge sector of the theory, see, e.g., Refs.~\cite{Bazavov:2007zz,Fischer:2007pf}. However,
{the} extension} of the present low-energy model to a so-called Polyakov-loop extended low-energy
model, see, e.g., Refs.~\cite{Meisinger:1995ih,Pisarski:2000eq,Mocsy:2003qw,Fukushima:2003fw,%
  Megias:2004hj,Ratti:2005jh, Roessner:2006xn,Sasaki:2006ww,Schaefer:2007pw,
  Schaefer:2009ui,Mizher:2010zb,Skokov:2010wb,Herbst:2010rf,Skokov:2010uh,Herbst:2013ail},
appears to be a first natural step towards a fully dynamical RG study of the QCD phase diagram in a finite volume.
For a first mean-field study of finite-volume effects with a Polyakov-loop extended low-energy QCD model, we refer the reader to Ref.~\cite{Bhattacharyya:2012rp}.
The corresponding RG flow equation, {which also takes  effects of meson fluctuations into account,} can be found
in Ref.~\cite{Braun:2011iz}. {In addition to the inclusion of confinement effects, it might be worthwhile to study the effect of
further bosonic composites, such as diquarks, on the finite-volume shift of the CEP since it has been found in infinite-volume studies that
they may affect the structure of the phase diagram and the location of the CEP significantly, see, e.g., Refs.~\cite{Roessner:2006xn,Strodthoff:2011tz,Strodthoff:2013cua}.}

In summary, our present investigation already shows that there is a qualitative effect of a finite
volume on the structure of the chiral phase diagram, {which can be} measured in terms of the coordinates of the 
CEP. The observed shift of the CEP in a finite volume could be useful to further guide 
present and future studies of the QCD phase diagram with lattice simulations as well as 
the experimental search for the CEP, and therefore it {may help} us to better understand the dynamics underlying strongly-interacting matter.

\acknowledgments 
The authors thank C.~S.~Fischer, H.~Gies, J.~M.~Pawlowski, L. von Smekal, and A.~Wipf
for useful discussions. {This work is supported by
  the DFG research training group GRK 1523/1, the Helmholtz
  International Center for FAIR within the LOEWE program of the State
  of Hesse, the DFG research cluster ``Structure and Origin of the
  Universe", the FWF grant P24780-N27 and by CompStar, a research
  networking programme of the European Science Foundation.}
\vfill

\bibliography{qcd}
\end{document}